\documentclass[aps,pra,twocolumn,amsmath,amssymb,showpacs,floatfix,superscriptaddress]{revtex4-1}

\usepackage{times}

\usepackage[utf8x]{inputenc}
\usepackage[english]{babel}
\usepackage[T1]{fontenc}
\usepackage{lmodern}
\usepackage{amssymb}
\usepackage{amsmath}
\usepackage{amsthm}
\usepackage[pdftex]{graphicx}
\usepackage{epstopdf}
\usepackage{braket}
\usepackage[bottom]{footmisc}

\usepackage{dcolumn}
\usepackage{bm}
\usepackage{color}
\usepackage{relsize,dsfont,mathrsfs,empheq,verbatim,upgreek}
\usepackage{etoolbox}
\usepackage[caption = false]{subfig}

\newtheorem{mylem}{Lemma}

\newcommand{\mathd}{\mathrm{d}}
\newcommand{\Tr}{\mathrm{Tr}}
\newcommand{\Lt}{\mathcal{L}}
\newcommand{\Ht}{\mathcal{H}}
\newcommand{\Dt}{\mathcal{D}}
\newcommand{\hilbert}{\mathscr{H}}

\begin{document}


\title{Open System Approach to Nonequilibrium Quantum Thermodynamics at Arbitrary Coupling}


\author{Alessandra Colla}

\affiliation{Institute of Physics, University of Freiburg, 
Hermann-Herder-Stra{\ss}e 3, D-79104 Freiburg, Germany}

\author{Heinz-Peter Breuer}

\affiliation{Institute of Physics, University of Freiburg, 
Hermann-Herder-Stra{\ss}e 3, D-79104 Freiburg, Germany}

\affiliation{EUCOR Centre for Quantum Science and Quantum Computing,
University of Freiburg, Hermann-Herder-Stra{\ss}e 3, D-79104 Freiburg, Germany}

\begin{abstract}
We develop a general theory describing the thermodynamical behavior of open quantum
systems coupled to thermal baths beyond perturbation theory. Our approach is based on
the exact time-local quantum master equation for the reduced open system states, and on
a principle of minimal dissipation. This principle leads to a unique prescription for the 
decomposition of the master equation into a Hamiltonian part representing coherent time 
evolution and a dissipator part describing dissipation and decoherence. Employing this
decomposition we demonstrate how to define work, heat and entropy production, formulate
the first and second law of thermodynamics, and establish the connection between violations 
of the second law and quantum non-Markovianity.
\end{abstract}

\date{\today}

\maketitle

\section{Introduction}\label{sec:Intro}

Quantum thermodynamics is concerned with the basic laws of equilibrium and nonequilibrium
thermodynamics in the quantum regime \cite{Gemmer2004,Schaller2014,Binder2018,Deffner2019,Landi2021}. 
One of the topical fundamental problems in this field is a unique and consistent definition 
of work, heat and entropy production for nonequilibrium processes in open quantum systems \cite{Breuer2002}
coupled to thermal reservoirs. Despite numerous proposals, a satisfactory and generally accepted definition 
of these quantities has not yet been advanced, in particular in the regime of strong system-reservoir interactions, 
and the topic remains highly controversial (see, e.g., 
Refs.~\cite{Weimer2008,Esposito2010,Teifel2011,Alipour2016,Strasberg2017,Campell2018,Strasberg2019,Rivas2020,Alipour2021,Landi2021}).
The case of equilibrium thermodynamics, which is presently dominated by the formalism of the so-called Hamiltonian of 
mean force \cite{Seifert2016} even at a quantum level \cite{Campisi2009},  also points at uncertainties in the definitions of 
such quantities in the strong coupling regime \cite{Talkner2016,Talkner2020}, with discussions still emerging 
\cite{Strasberg2020,Strasberg2020comment}. Moreover,  while a straightforward extension of the formalism to some 
non-equilibrium situations has been put forward \cite{Strasberg2020},  its legitimacy has already been contested 
\cite{Talkner2020comment}, and the matter seems far from resolved.

For weak couplings between the open system and the environmental baths one can often model 
the behavior of the open system by a quantum Markov process which leads 
to physically well-founded definitions of thermodynamic quantities and to corresponding 
formulations of the first and second law of quantum thermodynamics \cite{Spohn1978,Lebowitz1978,Kosloff2013}. 
However, for strong system-environment couplings, structured environmental spectral densities or low temperatures 
these definitions are no longer legitimized, as non-Markovian dynamics and memory effects become relevant 
\cite{Breuer2016a,deVega2017} and the treatment through a Markovian master equation fails. Above all, the absence of a 
unique, physically justified prescription for the treatment of the interaction energy between system and heat baths, 
which is non-negligible in this context, leads to ambiguities in the definitions of thermodynamic quantities 
\cite{Alipour2016,Seifert2016}.

Here, we propose an alternative strategy to develop a general theory for the quantum thermodynamics 
of open systems, which is valid for arbitrary system-environment couplings, temperatures and driving fields. Our
approach is based on the exact time-convolutionless quantum non-Markovian master equation for the open system, 
and on a general method of quantifying dissipation recently proposed by Hayden and Sorce \cite{Hayden2021}.
This method leads to a decomposition of the quantum master equation into coherent (reversible) and 
dissipative (irreversible) motion which is uniquely determined by a minimal size of the dissipator, 
which we therefore refer to as {\textit{principle of minimal dissipation}}. 
The application of this principle allows us to identify uniquely 
the contributions describing work and heat, to define entropy production and to formulate a first and second law 
of quantum thermodynamics. In addition we also discuss the connection between non-Markovian dynamics of the
open system and the emergence of negative entropy production rates, a topical issue which has attracted 
a lot of interest recently \cite{Alipour2017,Campell2018,Strasberg2019,Rivas2020}.

The paper is organized as follows. In Sec.~\ref{sec:eme} we recapitulate the basic features of the
description of open quantum systems by means of exact time-local master equations. The principle of minimal
dissipation is introduced and discussed in Sec.~\ref{sec:pmd}, which then enables us to define  in Sec.~\ref{sec:fslaws}
all relevant thermodynamics quantities, namely internal energy, work, heat and entropy production rate, and to
formulate a first and second law. Here, we also discuss the relation between quantum non-Markovianity and
negative entropy production rates. As a simple illustrative example we discuss in Sec.~\ref{sec:example} 
a two-state system interacting with a bosonic mode initially in a thermal equilibrium state.
Finally, in Sec.~\ref{sec:concl} we draw our conclusions and indicate directions of further research.
The appendix contains all relevant mathematical details about the invariance transformations of the generator 
of the master equation and the principle of minimal dissipation.

\section{Exact master equations}\label{sec:eme}
We consider an open quantum system $S$ which is coupled to an environment $E$ 
representing a heat bath initially in a thermal equilibrium state at temperature $T$. Let $\Phi_t$ be the quantum
dynamical map which propagates the open system's initial states $\rho_S(0)$ at time $t=0$ to
the corresponding states at time $t\geq 0$, i.e. $\rho_S(t) = \Phi_t [\rho_S(0)]$. For technical simplicity
we assume in the following that the Hilbert space $\hilbert$ of the open system is finite dimensional and 
that the initial states of the total system $S+E$ are given by a tensor product
$\rho_{SE}(0)=\rho_S(0)\otimes\rho_E(0)$, where $\rho_E(0)$ is a fixed thermal equilibrium (Gibbs) state 
of temperature $T$. It is well known that in this case the dynamical maps $\Phi_t$ represent a family of
completely positive and trace preserving (CPT) maps \cite{Breuer2002}. 
We assume that the total system $S+E$ is closed and governed by a Hamiltonian of the general form
\begin{equation} \label{Ham-total}
 H(t) = H_S(t) + H_E + H_I(t),
\end{equation}
where $H_S$ and $H_E$ are the free Hamiltonians of system and environment, respectively, and $H_I$ is the 
interaction Hamiltonian. Note that system and interaction Hamiltonian are allowed
to depend explicitly on time to include, e.g., an external driving or a turning on and off of the 
system-environment interaction.

The general evolution of the reduced density matrix can be described through an exact time-convolutionless (TCL) 
master equation:
\begin{equation} \label{tcl-master}
 \frac{d}{dt}\rho_S(t) = \Lt_t[\rho_S(t)],
\end{equation}
where the generator is related to the dynamical map by means of $\Lt_t=\dot{\Phi}_t \Phi_t^{-1}$.
We remark that the existence of the inverse of the dynamical map 
is a very weak assumption which may be assumed to hold in the generic case 
\cite{Stelmachovic2001a,Breuer2012a,Andersson2014a}.
From the requirement of Hermiticity and trace preservation, analogously to the treatment in \cite{Gorini1976a}, one finds 
that the generator has the following general structure \cite{Breuer2012a,Andersson2014a},
\begin{equation} \label{canonical}
 \Lt_t = {\mathcal{H}}_t + {\mathcal{D}}_t,
\end{equation}
where
\begin{equation} \label{Ham-part}
 {\mathcal{H}}_t [\rho_S] = -\mathrm{i} \left[K_S(t),\rho_S\right]
\end{equation}
represents a Hamiltonian part, given by the commutator of the density matrix with some effective
system Hamiltonian $K_S(t)$, and
\begin{equation} \label{Diss-part}
 {\mathcal{D}}_t [\rho_S] = \sum_{k}\gamma_{k}(t)\Big[L_{k}(t)
  \rho_S L_{k}^{\dag}(t) - \frac{1}{2}\big\{L_{k}^{\dag}(t)L_{k}(t),\rho_S\big\}\Big]
\end{equation}
is a so-called dissipator involving a set of generally time dependent rates $\gamma_k(t)$ and
Lindblad operators $L_k(t)$. The master equation \eqref{tcl-master} describes the full non-Markovian
quantum dynamics of open systems \cite{Breuer2016a} and, hence, all kinds of memory effects 
although there is no time-convolution over a memory kernel as, e.g., in the Nakajima-Zwanzig equation 
\cite{Nakajima1958,Zwanzig1960}. Master equations of this time-local form can be derived from the
microscopic Hamiltonian \eqref{Ham-total} by means of the time-convolutionless projection operator 
technique \cite{Shibata1977,Shibata1979}. If the decoherence rates $\gamma_k$, the Lindblad operators
$L_k$ and the effective system Hamiltonian $K_S$ are time independent the master equation \eqref{tcl-master}
obviously reduces to a master equation in the Gorini, Kossakowski, Sudarshan, Lindblad form
\cite{Gorini1976a,Lindblad1976} provided the $\gamma_k$ are positive. 
However, in general the rates $\gamma_k(t)$ can become negative without violating the
complete positivity of the dynamical map $\Phi_t$ \cite{Breuer2002,Breuer2016a}. Many exact master
equations of the time-local form \eqref{tcl-master} are known in the literature, such as the Hu-Paz-Zhang 
master equation for quantum Brownian motion \cite{Hu1992,Feraldi2016}, and the master equations for 
noninteracting bosons (fermions) linearly coupled to bosonic (fermionic) environments \cite{Zhang2012,Huang2020},
for specific spin-boson models \cite{Breuer1999b}, for general pure decoherence models \cite{Ban2010,Doll2008}, 
and for certain spin bath models \cite{Bhattacharya2017}.

Our first goal is to identify $K_S(t)$ as the operator associated to the physical effective energy of the system, and from this 
construct exact thermodynamic quantities. The problem with this ambition is that the decomposition \eqref{canonical}
of the generator $\Lt_t$ of the master equation is highly non-unique. In fact, if one performs for each fixed time $t$
the transformation
\begin{eqnarray}
 L_k &\longrightarrow & L_k - \alpha_k \mathbb{I} \label{trafo-1} \\
 K_S &\longrightarrow & K_S + \sum_{k} \frac{\gamma_k}{2\mathrm{i}}
 \left( \alpha_k L^{\dag}_{k} - \alpha^*_k L_k \right) + \beta \mathbb{I} \label{trafo-2}
\end{eqnarray}
induced by arbitrary time dependent scalar functions $\{\alpha_k(t)\}$ and $\beta(t)$, the generator $\Lt_t$ remains 
invariant, while the Hamiltonian part ${\mathcal{H}}_t$, i.e. the effective Hamiltonian $K_S(t)$, and the dissipator 
${\mathcal{D}}_t$ do change in general in a nontrivial way 
\footnote{There is an additional invariance induced by an indefinite unitary transformation of the set 
of Lindblad operators; since this is not particularly relevant for the topic of this paper, we report it 
for completeness in Appendix \ref{app-A}.}.
If one is to make use of the two contributions separately, one must find a physical 
motivation for the particular choice of splitting used, as the derived physical quantities will have different shapes and values 
depending on this choice. In the following we review the method developed in Ref.~\cite{Hayden2021}, which leads to a
unique splitting of the generator into Hamiltonian part and dissipator appropriate for our purpose.

\section{Principle of minimal dissipation}\label{sec:pmd}
As mentioned, the form of the generator of the master equation
given by Eqs.~\eqref{canonical}-\eqref{Diss-part} is a 
consequence of Hermiticity and trace preservation. Let $\mathfrak{S}(\hilbert)$ denote the space of linear maps 
$\Lt : \mathcal{B}(\hilbert) \rightarrow \mathcal{B}(\hilbert)$, i.e. the space of superoperators of the open system.
Then one can introduce a real vector space $\mathfrak{htp}(\hilbert)$ consisting of all superoperators
$\Lt \in \mathfrak{S}(\hilbert)$ which are Hermiticity and trace preserving, i.e., which satisfy the conditions
\begin{eqnarray}
 \Lt [A^{\dagger}] = \Lt [A]^{\dagger}, \quad \Tr \big\{\Lt [A]\big\} = 0 \quad  \forall A \in \mathcal{B}(\hilbert).
\end{eqnarray}
Generators of quantum master equations are indeed all elements of $\mathfrak{htp}(\hilbert)$, although not every 
element in $\mathfrak{htp}(\hilbert)$ is the generator of a completely positive evolution (or even positive, for that matter). 

In order to quantify the size of superoperators one introduces an appropriate norm on the space $\mathfrak{htp}(\hilbert)$
by means of the definition 
\begin{equation} \label{super-norm}
 || \Lt ||^2 = \overline{\bra{\psi} \overline{\Lt [\ket{\phi}\bra{\phi}]^2}\ket{\psi}}
\end{equation}
proposed recently in Ref.~\cite{Hayden2021}. A nice feature of this norm is the fact that it is induced by the scalar product 
\begin{equation}\label{scalarprod}
 \langle \mathcal{L}_1 , \mathcal{L}_2 \rangle 
 = \overline{\bra{\psi} \overline{\mathcal{L}_1\big[\ket{\phi}\bra{\phi}\big] \mathcal{L}_2\big[\ket{\phi}\bra{\phi}\big]}\ket{\psi}},
\end{equation}
where $\mathcal{L}_1$, $\mathcal{L}_2 \in \mathfrak{htp}(\hilbert)$, which means that one has
$|| \Lt ||^2=\langle \mathcal{L},\mathcal{L}\rangle$. In these definitions $\ket{\phi}$ and $\ket{\psi}$ are
normalized random state vectors and the overline denotes the corresponding average over the Haar measure on the
unitary group \cite{Collins2006}. 
As argued in \cite{Hayden2021} the norm $|| \Lt ||$ represents a 
measure for the average size of the superoperator $\Lt$. This measure is obtained by acting with $\Lt$ on a random
input state $\ket{\phi}\bra{\phi}$, squaring the result, taking the expectation value with respect to another
independent random state $\ket{\psi}\bra{\psi}$, averaging and, finally, taking the square root. Note that within the
averaging procedure all pure states have equal weight because $\ket{\phi}$ and $\ket{\psi}$ are distributed
according to the Haar measure.

\begin{figure}[tp]
\includegraphics[width=0.8\columnwidth]{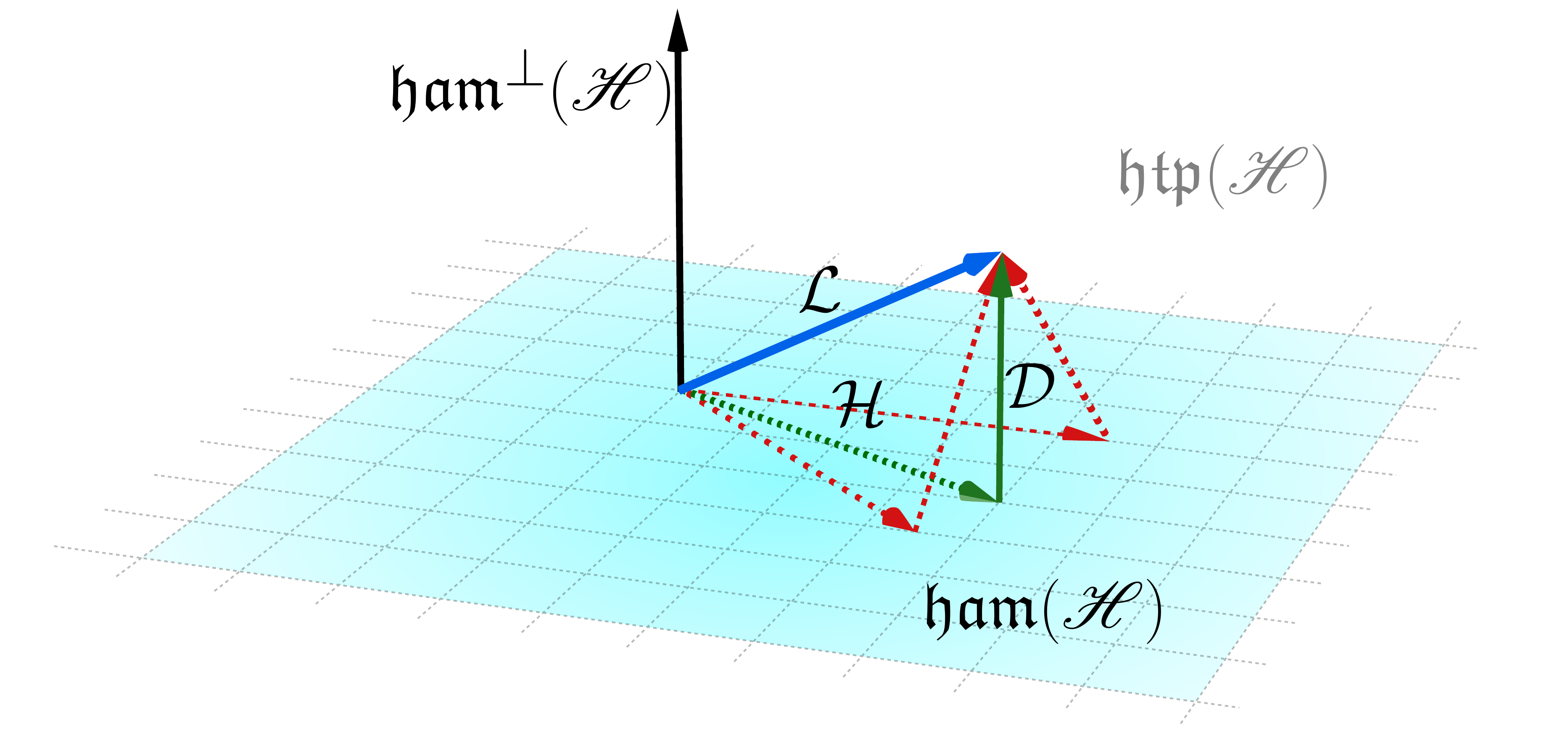}
\caption{Decomposition of the generator $\Lt$ into some Hamiltonian contribution $\Ht \in \mathfrak{ham}(\hilbert)$ 
and a dissipator $\Dt$. The dissipator part $\Dt$ can generically admit a component on $\mathfrak{ham}(\hilbert)$; 
the unique dissipator given by the orthogonal decomposition \eqref{decomp}, defined through the scalar product 
\eqref{scalarprod}, is the one with minimal associated norm $|| \Dt ||$.}
\label{figure1}
\end{figure}

We can now give a simple geometric picture of the principle of minimal dissipation (see Fig.~\ref{figure1}). To this end,
let us define the subspace $\mathfrak{ham}(\hilbert)$ of $\mathfrak{htp}(\hilbert)$ which is formed by all Hamiltonian
superoperators of the form of $\mathcal{H}_t$ (see Eq.~\eqref{Ham-part}), i.e. by all superoperators
$\mathcal{H} \in \mathfrak{htp}(\hilbert)$ for which there exists a Hermitian operator $H$ such that
$\mathcal{H} [A] = - \mathrm{i} [ H, A] \;\; \forall A \in \mathcal{B}(\hilbert)$. Given a superoperator $\Lt_t$ one
defines the Hamiltonian part $\mathcal{H}_t$ as the orthogonal projection of $\Lt_t$ onto the subspace 
$\mathfrak{ham}(\hilbert)$ defined by the scalar product \eqref{scalarprod}. Denoting the corresponding projection
by $\Pi$ one thus has
\begin{equation} \label{decomp}
\mathcal{H}_t = \Pi(\Lt_t), \quad \mathcal{D}_t = \Lt_t - \Pi(\Lt_t).
\end{equation}
Note that this splitting corresponds to the orthogonal decomposition 
$\mathfrak{htp}(\hilbert)=\mathfrak{ham}(\hilbert) \oplus \mathfrak{ham}^{\perp}(\hilbert)$ with respect to the scalar 
product associated to the norm. 

It is clear that the above construction yields unique expressions for the Hamiltonian part
and the dissipator of the master equation. Moreover, among all possible decompositions of the generator into
a Hamiltonian part and a dissipator, the one given by Eq.~\eqref{decomp} has the least norm of the dissipator.
This is why we call this the principle of minimal dissipation: it states that we uniquely define the dissipator of the
master equation, which will be associated with heat exchange, by making its average effect on physical states
as small as possible, shifting as much as possible into the Hamiltonian contribution. 
In view of this interpretation the choice of the particular norm \eqref{super-norm} is physically
reasonable because it is a democratic norm which measures the size of superoperators 
by giving equal weight to all pure states.
Most remarkable,
it has been demonstrated in Ref.~\cite{Hayden2021} that a minimal norm of the dissipator exactly corresponds to taking 
traceless Lindblad operators (for an alternative proof see Appendix \ref{app-B}). 
Thus, to satisfy the principle of minimal dissipation in practice one only has to ensure that the Lindblad operators are 
traceless for all times, which can always be achieved by a suitable transformation of the form \eqref{trafo-1}. We note that 
the Lindblad operators are often already traceless when the master equation is derived perturbatively from a 
microscopic model. This is the case, for example, for the standard weak-coupling semigroup master equation 
in Lindblad form \cite{Breuer2002} whose Lindblad operators are eigenoperators of the system Hamiltonian $H_S$ and,
hence, are automatically traceless.

\section{First and second law of thermodynamics}\label{sec:fslaws}
Fixing the unique splitting of the generator into $\mathcal{H}_t$ 
and $\mathcal{D}_t$ according to the principle of minimal dissipation allows us to identify the associated Hamiltonian 
$K_S(t)$ as an effective Hamiltonian for the system, that encompasses the collective effects of the bath onto it. 
The variation of internal energy of the system can be then defined as 
$\Delta U_S (t) = \Tr \{K_S(t)\rho_S(t)\} - \Tr \{K_S(0)\rho_S(0)\}$ such that the first law arises naturally as
\begin{eqnarray}\label{first-law}
\Delta U_S (t) = \delta W_S(t) + \delta Q_S(t)
\end{eqnarray}
by defining work and heat contributions as
\begin{eqnarray}
 \delta W_S(t) &=& \int_0^t \mathd \tau \, \Tr \big\{ \dot{K}_S(\tau) \rho_S(\tau) \big\},  \label{work} \\
 \delta Q_S(t) &=& \int_0^t \mathd \tau \, \Tr \big\{ K_S(\tau) \dot{\rho}_S(\tau)  \big\}.  \label{heat}
\end{eqnarray}
Here, the heat contribution turns out to be only due to the dissipative part of the evolution of the density matrix 
since the above can be written as
\begin{equation}
 \delta Q_S(t) = \int_0^t \mathd \tau \, \Tr \big\{ K_S(\tau) \mathcal{D}_{\tau}[\rho_S(\tau)] \big\}.
\end{equation}

It is worth mentioning that it is possible for the effective Hamiltonian $K_S$ to be time dependent even if the original system 
Hamiltonian $H_S$ is not. For instance, this is the case for the example discussed in Sec.~\ref{sec:example}. 
Through this our approach admits the appearance of effective work done on the system as a result of the interaction with 
the bath. This feature is shared with other approaches (see, e.g., \cite{Weimer2008,Teifel2011,Alipour2016}). 
Note that this does not contradict the idea that a change in internal energy of a closed system should be identified with 
work. It is a well-known fact, which holds of course also in our formalism, that a change in the internal energy of the total 
system can only be due to explicit time dependencies of the total system Hamiltonian. However, contrary to what is 
assumed in many alternative approaches \cite{Rivas2020,Strasberg2020,Talkner2020}, this does not imply that the change 
in internal energy of the total system should be identified with work done \textit{on the open system only}.  
As emerges from our formalism the environment can perform work on the open system, even when the total
system Hamiltonian is time independent and, hence, the internal energy of the total closed system does not change:
In such a case there is an exchange of energy between the open system and its environment which manifests itself in the
time dependence of the effective system Hamiltonian $K_S(t)$ and must be interpreted as mechanical work.

As a result of the above definition of heat exchange, the entropy production is defined as
\begin{eqnarray}
 \Sigma_S (t) = \Delta S_S(t) - \beta \delta Q_S(t) \;,
\end{eqnarray}
with $\Delta S_S(t)=S(\rho_S(t))-S(\rho_S(0))$ the change of the von Neumann entropy of the reduced system and 
$\beta=1/\mathrm{k_B}T$ the inverse temperature of the bath. 
For simplicity we assume here that the environment is sufficiently large such that its temperature 
can be regarded as effectively constant as has been discussed recently \cite{Strasberg2021}.
An alternative expression for the entropy production is given by
\begin{eqnarray}\nonumber
 \Sigma_S (t) &=&  S(\rho_S(0)|| \rho_{S}^{\mathrm{G}}(0)) - S(\rho_S(t)||\rho_S^{\mathrm{G}}(t)) \\ 
 & & -\int_0^t \mathd \tau \, \Tr \big\{ \rho_S(\tau)\partial_\tau \ln \rho_S^{\mathrm{G}}(\tau) \big\},  \label{DLutz}
\end{eqnarray}
with $S(\rho_A || \rho_B )$ the relative entropy of the states $\rho_A$ and $\rho_B$, and where one utilizes the 
instantaneous Gibbs states associated to the effective Hamiltonian $K_S(t)$, namely 
$\rho_S^{\text{G}}(t)= e^{- \beta K_S(t) }/Z_S(t)$. It is important to note that the structure of expression \eqref{DLutz}
for the entropy production is the same as the one derived in Ref.~\cite{Deffner2011}. The crucial difference is,
however, that in our expression the effective Hamiltonian $K_S(t)$ appears, while in \cite{Deffner2011} this
Hamiltonian is replaced by the microscopic system Hamiltonian $H_S(t)$ (see Eq.~\eqref{Ham-total}).
As a consequence of the fact that $K_S(t)$ contains the effective influence of the bath on the open system, 
our expression is valid also outside of the weak-coupling regime. As expected, in the limit of vanishing coupling 
our expression \eqref{DLutz} for the entropy production reduces for arbitrary driving to the one obtained in 
\cite{Deffner2011}. To see this we recall that the master equation \eqref{tcl-master} and, in particular, the
effective Hamiltonian $K_S(t)$ can be derived from the total Hamiltonian \eqref{Ham-total} by means of the 
time-convolutionless projection operator technique \cite{Shibata1977,Shibata1979,Breuer2002}. This technique
leads to a perturbation expansion for the effective Hamiltonian $K_S(t,\lambda)$ in powers of the size 
$\lambda$ of the system-environment interaction which takes the following form,
\begin{equation} \label{weak-coupling-limit}
 K_S(t,\lambda) = H_S(t) + \lambda^2 G_S(t) + {\mathcal{O}}(\lambda^4).
\end{equation}
This shows that in the limit $\lambda \to 0$ the effective Hamiltonian reduces to the bare Hamiltonian $H_S(t)$ 
of the open system, appearing in the microscopic Hamiltonian \eqref{Ham-total} of the total system. Consequently, 
in this limit also \eqref{DLutz} reduces to the expression derived in \cite{Deffner2011}. Moreover,
if $H_S$ is time independent and the open system dynamics is described by a quantum Markovian
semigroup we recover the earlier results of Ref.~\cite{Spohn1978}. 

Taking the time derivative of Eq.~\eqref{DLutz} we obtain the entropy production rate
\begin{eqnarray} \label{entr-prod-rate}
 \sigma_S(t) &\equiv& \dot{\Sigma}_S(t) 
 = - \frac{\mathd}{\mathd \tau} \Bigg|_{\tau=0} S(\rho_S(t+\tau)|| \rho_S^{\mathrm{G}}(t)) \nonumber \\
 &=&  - \Tr \big\{ \mathcal{D}_t [\rho_S(t)] \big( \ln \rho_S(t) - \ln \rho_S^{\mathrm{G}}(t) \big) \big\} \;.
\end{eqnarray}
The expression given in the first line relates the entropy production rate to the derivative of the relative entropy,
where the notation indicates that $\rho_S^{\mathrm{G}}$ should be regarded as a constant under the time derivative,
while the second line connects it directly to the dissipator of the master equation.
Let us assume that the Gibbs state $\rho_S^{\text{G}}(t)$ represents an instantaneous fixed point of the 
evolution \cite{Strasberg2019,Altaner2017}, namely that there is no instantaneous dissipation for this state: 
$\Lt_t[\rho_S^{\text{G}}(t)]= \mathcal{D}_t[\rho_S^{\text{G}}(t)] = 0$. 
Under this condition one can show that the entropy production rate is positive
if the dynamical map $\Phi_t$ is P-divisible \cite{Breuer2016a}, i.e., one has
\begin{equation} \label{second-law}
 \sigma_S(t) \geq 0 \; ,
\end{equation}
which corresponds to the second law. 
Note that we associate here the second law with the positivity of the
entropy production rate, which implies an increase of entropy over all time intervals 
(see, e.g., Ref.~\cite{Strasberg2017}).
To prove \eqref{second-law} one uses the fact that the relative entropy
decreases under the application of a positive trace preserving map to both of its arguments \cite{Reeb2017}.

Finally, it might be interesting to discuss how possible violations of the second law are related to the
non-Markovianity of the underlying dynamics, i.e. to the presence of quantum memory effects \cite{Breuer2016a}. 
To this end, we employ the definition for quantum non-Markovianity based on the information flow between 
the open system and its environment \cite{Breuer2009b,Breuer2016a}. The key idea is to characterize Markovian
behavior in the quantum regime through a continuous loss of information, i.e. by a flow of information from
the open system to the environment. Correspondingly, quantum memory effects feature a flow of information
from the environment back to the system. Quantifying the information content by means of the distinguishability
of quantum states as measured by the Hellstrom matrix, one can show that Markovianity of quantum processes 
in open systems is equivalent to P-divisibility of the corresponding dynamical map 
\cite{Chruscinski2011a,Wissmann2015a}. Recall that P-divisibility means that the propagator $\Phi_{t,s}$
which maps the open system states at time $s$ to the open system states at time $t$
is a positive map for all $t \geq s \geq 0$. One the other hand, we have just seen that
under the condition that the Gibbs state is an instantaneous fixed point P-divisibility implies positivity
of the entropy production rate. We conclude that in order for Eq.~\eqref{second-law} to be violated the process 
must break P-divisibility and, hence, must be non-Markovian. Thus, we see that non-Markovianity, i.e. 
memory effects are a necessary condition for violations of the second law.

\section{Example}\label{sec:example}

To illustrate our theory we briefly discuss the model of a two-state atom, regarded as the open system, 
coupled to a single harmonic oscillator mode, which is also known as Jaynes-Cummings model. 
The total Hamiltonian is time independent and of the form of Eq.~\eqref{Ham-total}. The time independent system 
Hamiltonian is given by
\begin{equation} \label{H_S_JC}
 H_S = \omega_0 \sigma_+\sigma_-,
\end{equation}
where $\omega_0$ denotes the transition frequency of the two-state system with excited state $|1\rangle$ and
ground state $|0\rangle$, while $\sigma_\pm$ are the usual Pauli raising and lowering operators.
The environmental Hamiltonian is given by $H_E = \omega b^\dagger b$, where $\omega$ is the 
eigenfrequency of the harmonic oscillator and $b$, $b^\dagger$ denote the annihilation and creation operator, respectively. 
Finally, the time independent interaction Hamiltonian is taken to be of the Jaynes-Cummings form 
\begin{equation}
 H_I=g(\sigma_+b + \sigma_-b^\dagger),
\end{equation}
where $g$ is a coupling constant. 

\begin{figure}[htp]
\includegraphics[width=1.0\columnwidth]{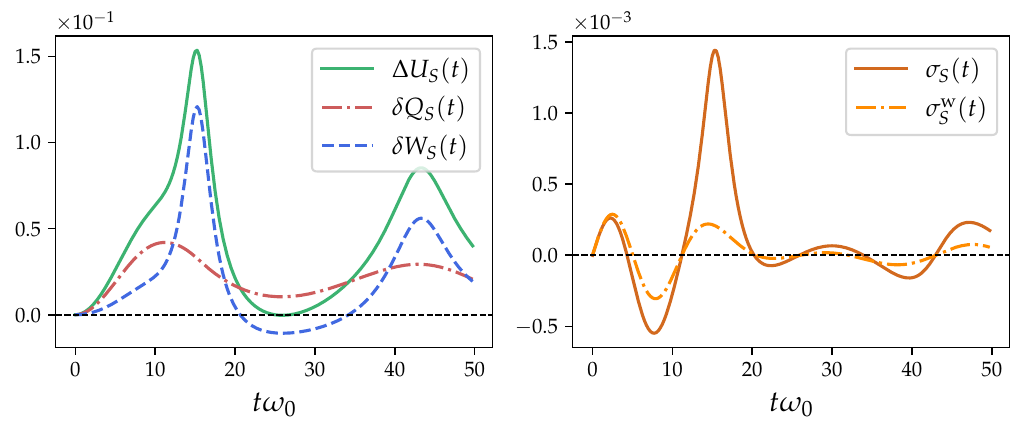}
\caption{Left: Change of internal energy $\Delta U_S$, work $\delta W_S$ and heat $\delta Q_S$ 
as a function of time for the Jaynes-Cummings model with effective system Hamiltonian \eqref{K_S_JC}.
Right: Comparison between the entropy production rate $\sigma_S$ according to Eq.~\eqref{entr-prod-rate}
and its weak-coupling version $\sigma_S^\mathrm{w}$ for the same model. Parameters: $g=0.1\omega_0$, 
$\omega = 0.9\omega_0$, $\mathrm{k_B}T=\omega_0$, $\rho_S^{11}(0)=0.25$ and  $\rho_S^{10}(0)=0$.}
\label{figure2}
\end{figure}

Assuming the oscillator to be initially in a thermal equilibrium state at a certain temperature $T$
one can derive an exact master equation for the dynamics of the two-state system \cite{Smirne2010} 
which is of the form given by Eqs.~\eqref{tcl-master}-\eqref{Diss-part}. The dissipator of the master equation
can be written in terms of the three traceless and time-independent Lindblad operators $L_1=\sigma_-$,
$L_2=\sigma_+$ and $L_3=\sigma_z$ with corresponding time dependent rates $\gamma_k(t)$, $k=1,2,3$.
The effective system Hamiltonian takes the form 
\begin{equation} \label{K_S_JC}
 K_S(t) = \big[ \omega_0+\Delta\omega(t) \big] \sigma_+\sigma_-.
\end{equation}
Comparing Eqs.~\eqref{H_S_JC} and \eqref{K_S_JC} we see that the interaction of the system with the
environmental mode leads to a time dependent frequency shift $\Delta\omega(t)$, which in general
depends on the system-environment coupling $g$ and on temperature $T$.
Further details and the expression for the frequency renormalization $\Delta\omega(t)$ may be found 
in  \cite{Smirne2010}. Here, we plot the results for the decomposition \eqref{first-law} of the change of
internal energy into work \eqref{work} and heat \eqref{heat}, and the comparison between the entropy production rate 
\eqref{entr-prod-rate} and its weak-coupling counterpart \cite{Spohn1978}, see Fig.~\ref{figure2}. 
We observe that there is a significant contribution of work to the change of internal energy, although the total
Hamiltonian is time independent, which is due to the time dependence of the 
effective system Hamiltonian \eqref{K_S_JC}. We also see that the magnitude of the entropy production rate
is substantially larger than the one given by the weak-coupling expression.

\section{Conclusions and remarks}\label{sec:concl}
We have formulated a general nonperturbative approach to the quantum 
thermodynamics of open systems based on the exact time-convolutionless master equation of the open system
and on the principle of minimal dissipation, which allows to develop unique expressions for work, heat and entropy 
production. As we have demonstrated the principle of minimal dissipation leads to a unique effective system Hamiltonian 
$K_S(t)$ which generates the coherent part of the open system dynamics and defines the internal energy of the system as 
it results from the cooperative effect of system and environment.
As we have seen the effective system Hamiltonian $K_S(t)$ can be time dependent even though the total microscopic 
system Hamiltonian $H$ in Eq.~\eqref{Ham-total} is time independent. Thus, for non-Markovian dynamics the environment 
can do work on the open system, or extract work from it, as is illustrated in the example of Sec.~\ref{sec:example}.
In our opinion this is an important consequence of our approach which could lead to interesting applications. 
We further emphasize that our strategy of constructing an effective open 
system Hamiltonian can also be applied to approximate time-local master equations, such as the quantum optical, the 
Brownian motion, the Redfield, or, more generally, the TCL master equations in any finite order of perturbation theory 
\cite{Breuer2002}. 

An important feature of our theory is the fact that it only refers to open system variables, i.e., only degrees of freedom 
of the open system enter the definitions of thermodynamic quantities. Thus, in contrast to other approaches 
(see, e.g., Ref.~\cite{Esposito2010}) the application of our method does not require the control or measurement of 
environmental variables. As a consequence we were able to establish under certain conditions a connection of our 
formulation of the second law with the concept of quantum non-Markovianity based on the information flow between system 
and environment. This connection yields a natural information theoretic interpretation because it implies that a
backflow of information from the environment to the open system is a necessary condition for violations of the
second law. 

We finally remark that the pillars of our approach, namely the TCL master equation and the principle of minimal 
dissipation, have been formally proven only for open systems with finite-dimensional Hilbert spaces. The 
generalization of the method developed here to the case of (bounded or even unbounded) generators on an 
infinite-dimensional Hilbert space represents a challenging mathematical project, but is extremely interesting 
and relevant from a physical point of view.

\acknowledgments

We thank Bassano Vacchini for many fruitful discussion and helpful comments.
This project has received funding from the European Union's Framework Programme for Research and
Innovation Horizon 2020 (2014-2020) under the Marie Sk\l{}odowska-Curie Grant Agreement No.~847471.


\appendix


\section{Invariance of the generator}\label{app-A}

The generator $\Lt = - \text{i} [ K, \cdot ] + \mathcal{D}$ is invariant under the addition of a time dependent constant to the Lindblad operators, and a consequent additional term to the Hamiltonian $K$, although this is not the most general invariance. In fact, it is also permitted to modify the dissipator by absorbing the rate amplitudes into the Lindblad operators, or to mix them through a generalized unitary transformation which preserves the sign of the rates. 
To see this, let us first rewrite the dissipator at a fixed time $t$, in particular by grouping the Lindblad operators associated to positive rates to lower indices, such that we can express it in the following way:
\begin{equation}
\mathcal{D} = \sum_{ij}J_{ij} \sqrt{|\gamma_i||\gamma_j|} \left[ L_i \rho L_j^{\dagger} - \frac{1}{2 }\{  L_i^{\dagger} L_j, \rho\} \right] \; ,
\end{equation}
where we defined the matrix describing the sign of the rates
\begin{equation}
J = \begin{bmatrix}
\mathbb{I}_p & 0\\
0 & - \mathbb{I}_q
\end{bmatrix} \; ,
\end{equation}
and where $p$ ($q$) is taken to be the number of positive (negative) rates. Note that this exact division into positive and negative rates may depend on time; still, this description is applicable at any fixed time $t$.
Defining the bilinear form
\begin{equation}
\left( \vec{v} , \vec{w}\right) := \vec{v} \cdot J \vec{w}
\end{equation}
and absorbing the rates $\sqrt{|\gamma_i|}$ into the definition of the Lindblad operators, we can formally rewrite the dissipator as
\begin{equation}
\mathcal{D} = (\vec{L}, \rho \vec{L}^{\dagger}) - \frac{1}{2} \{ (\vec{L}^{\dagger}, \vec{L}) , \rho\} \; ,
\end{equation}
where we have defined $\vec{L}$ as the vector of all $N^2-1$ Lindblad operators.

It is straightforward to prove that the generator $\Lt$ is at each time invariant under the transformation
\begin{eqnarray}
&\vec{L} &\longrightarrow  \Upsilon \vec{L} + \vec{\alpha} \\
&K &\longrightarrow  K + \frac{1}{2 \text{i}} \left[ (\vec{\alpha}^*, \Upsilon \vec{L}) - (\vec{\alpha}, \Upsilon^* \vec{L}^{\dagger}) \right] + \beta
\end{eqnarray}
given by the parameters $(\vec{\alpha}\in \mathbb{C}^{N^2-1}, \Upsilon \in \mathrm{U}(p,q), \beta \in \mathbb{R})$, where $U(p,q)$ is the indefinite (sometimes also called generalized) unitary group of matrices satisfying
\begin{equation}
\Upsilon^{\dagger} J \Upsilon = J \; .
\end{equation}
The transformations of the generator obey the following group property
\begin{equation}
(\vec{\alpha}', \Upsilon' , \beta')(\vec{\alpha}, \Upsilon , \beta) = (\vec{\alpha}'+ \Upsilon' \vec{\alpha} , \Upsilon' \Upsilon , \beta' + \beta + \text{Im}(\vec{\alpha}', \Upsilon' \vec{\alpha})) \; .
\end{equation}
When the dynamics happens to be CP-divisible, namely when $J \equiv \mathbb{I}$, this invariance reduces to the one of a generator of a Lindblad master equation \cite{Breuer2002} where $\Upsilon$ is a unitary transformation, albeit still with time dependent parameters.

\section{Derivation of the dissipator of minimal norm}\label{app-B}

The space of generators of quantum master equations can be orthogonally decomposed into
\begin{equation}
\mathfrak{htp}(\hilbert) = \mathfrak{ham}(\hilbert)\oplus \mathfrak{ham}^{\perp}(\hilbert) \; ,
\end{equation}
such that at fixed time $t$, any generator $\Lt$ can be written as $\Lt = \Ht + \Dt $, with $\Ht \in \mathfrak{ham}(\hilbert)$. To find the unique decomposition such that $\Dt$ is in the orthogonal subspace $\mathfrak{ham}^{\perp}(\hilbert)$, we project the generator onto $\mathfrak{ham}(\hilbert)$:
\begin{equation}
\Ht = \Pi(\Lt) = \sum_j \Ht_j \langle \Ht_j , \Lt \rangle \; ,
\end{equation}
with $\{\Ht_j\}_{j=1}^{N^2-1}$ an orthonormal basis of $\mathfrak{ham}(\hilbert)$, through the scalar product on $\mathfrak{htp}(\hilbert)$
\begin{equation}\label{scalarprod-app}
\langle \mathcal{X}_1 , \mathcal{X}_2 \rangle = \overline{\bra{\psi} \overline{\mathcal{X}_1[\ket{\phi}\bra{\phi}] \mathcal{X}_2[\ket{\phi}\bra{\phi}]}\ket{\psi}} \; .
\end{equation}
Here, $\ket{\phi}$ and $\ket{\psi}$ are random Haar states
\begin{equation}
\ket{\phi}= U \ket{\phi_0} \; ,
\end{equation}
with $\ket{\phi_0}$ a fixed normalized state and $U \in \mathrm{U}(N)$ a random unitary, while $\overline{\bra{\phi} X \ket{\phi}}$ denotes the Haar average of an operator $X$ over $\ket{\phi}$.

As a recurring tool, let us notice that, as a consequence of Schur's Lemma and the invariance of the Haar measure under multiplication with unitaries, it holds for any operator $X \in \mathcal{B}(\hilbert)$ that
\begin{equation}
X' := \int \mathd \mu(U) U^{\dagger} X U = c \cdot \mathbb{I} \; , \quad c \in \mathbb{C} \;,
\end{equation}
so that, by taking the trace of the above, one finds
\begin{equation}\label{idtrace}
\int \mathd \mu(U) U^{\dagger} X U = \frac{\Tr \{ X\}}{N} \mathbb{I} \;.
\end{equation}
Naturally this entails that the Haar average of $X$ is simply given by
\begin{equation} \label{average}
\overline{\bra{\psi} X \ket{\psi}} = \frac{\Tr X}{N} \; ,
\end{equation}
and that the norm associated to the scalar product \eqref{scalarprod-app} reads:
\begin{equation}\label{norm}
|| \Lt ||^2 = \frac{1}{N} \Tr \left\{ \overline{\Lt \left[ \ket{\phi} \bra{\phi}\right]^2} \right\} \; .
\end{equation}

To find the expression for the Hamiltonian $K $ generating $\Ht = - \text{i} [K, \cdot]$, we exploit the fact that $\mathfrak{ham}(\hilbert)$ is isomorphic to the space of traceless hermitian operators $\mathrm{Herm_{tl}}(\hilbert) = \{ H \in \mathcal{B}(\hilbert) | H^{\dagger}=H, \; \Tr \{ H \} = 0 \}$. With this we can exploit the connection between a basis of superoperators in $\mathfrak{ham}(\hilbert)$ and one of operators in $\mathrm{Herm_{tl}}(\hilbert)$ found in the following Lemma.
\begin{mylem}\label{lemma:basis}
Each element of an orthonormal basis $\{\Ht_j\}_{j=1}^{N^2-1}$ of $\mathfrak{ham}(\hilbert)$ with respect to the scalar product \eqref{scalarprod-app} is such that 
\begin{equation}
\Ht_j= -\mathrm{i} [H_j, \cdot] \; ,
\end{equation}
where $\{H_j\}_{j=1}^{N^2-1}$ is an orthogonal basis with respect to the Hilbert-Schmidt product on $\mathrm{Herm_{tl}}(\hilbert)$ satisfying
\begin{equation}\label{orthoH}
\Tr \{ H_i H_j\} = \frac{N(N+1)}{2} \delta_{ij} \; .
\end{equation}
\begin{proof}
From \eqref{average} and the normalization of $\ket{\phi}$, one has that 
\begin{align} \nonumber
\langle \Ht_i , \Ht_j \rangle = & \frac{1}{N} \Tr \left\{ \overline{\Ht_i[\ket{\phi}\bra{\phi}] \Ht_j[\ket{\phi}\bra{\phi}]} \right\} \\ \nonumber
 = & \underbrace{\frac{1}{N} \Tr \left\{ \{H_i,H_j \} \overline{\ket{\phi}\bra{\phi}} \right\} }_{(a)}  \\
& \quad \quad \underbrace{ - \frac{1}{N} \Tr \left\{ 2 H_i \overline{ \ket{\phi} \bra{\phi} H_j \ket{\phi}\bra{\phi} } \right\} }_{(b)} \; .
\end{align}
The second moment term $(a)$ is simply given by $\frac{2}{{N^2}} \Tr \left\{ H_i H_j \right\}$ through identity \eqref{idtrace}, 
while the fourth moment Haar integral found in $(b)$ is less straightforward to calculate. One can make use of the following 
general formula (see, e.g., Ref.~\cite{Gessner2013}):
\begin{align}\label{fourthmoment} \nonumber
\int \mathd \mu (U) & U X_1 U^{\dagger} X_2 U X_3 U^{\dagger} = \\ \nonumber
 =& {N \Tr \{ X_3 X_1 \} - \Tr \{X_1\} \Tr \{ X_3\} \over N(N^2-1)} \left(\Tr \{ X_2 \} \right) \mathbb{I} \\
&+  {N \Tr \{X_1\} \Tr \{ X_3\}-  \Tr \{ X_3 X_1 \} \over N(N^2-1)} X_2  \; ,
\end{align}
and the tracelessness of $H_j$ to find that $(b)$ reads $ - {2\over {N^2(N+1)}} \Tr \left\{ H_i H_j \right\}$. Imposing the orthogonality of the basis of $\mathrm{Herm_{tl}}(\hilbert)$ \eqref{orthoH}, one recovers 
\begin{equation}
\langle \Ht_i , \Ht_j \rangle = \delta_{ij} \; .
\end{equation}
\end{proof}
\end{mylem}

It is then straightforward that the Hamiltonian operator associated to $\Ht$ is given by
\begin{equation}
K = \sum_j H_j \langle \Ht_j , \Lt \rangle \;.
\end{equation}
To find the expression for the coefficients $\langle \Ht_j , \Lt \rangle$, we make use of the pseudo-Kraus 
representation for the generator, which states that any Hermiticity preserving map can be written as
$\Lt [\rho] = \sum_k \gamma_k E_k \rho E_k^{\dagger}$ with some operators $E_k$ and real (not necessarily
positive) coefficients $\gamma_k$ \cite{Choi1975}. Employing \eqref{average} we then find
\begin{eqnarray} \nonumber
\langle \Ht_j , \Lt \rangle &= -{ \text{i} \over N}\sum_k \gamma_k \Tr  \Big\{ & \overline{[H_j, \ket{\phi}\bra{\phi}] E_k  \ket{\phi}\bra{\phi} E_k^{\dagger}} \Big\} \\ \nonumber
&= -{ \text{i} \over N}\sum_k \gamma_k \Tr \Big\{ &H_j \overline{ \ket{\phi}\bra{\phi} E_k \ket{\phi}\bra{\phi}} E_k^{\dagger}  \\ \nonumber
& & \quad- H_j E_k \overline{ \ket{\phi}\bra{\phi} E_k^{\dagger} \ket{\phi}\bra{\phi}} \Big\}  \\
&= -{1 \over N^2 (N+1) } \sum_k  & \gamma_k \Tr \{ H_j Y_k \} \; ,
\end{eqnarray}
where we used again expression \eqref{fourthmoment} for the fourth moment Haar integral and have defined the operators 
\begin{equation}
Y_k = \text{i} (\Tr \{E_k\} E_k^{\dagger} - \Tr \{E_k^{\dagger}\} E_k )\; .
\end{equation}
From the completeness of the basis $H_j$ and Lemma \ref{lemma:basis}, and since each operator $Y_k$ is in $\mathrm{Herm_{tl}}(\hilbert)$, one has
\begin{equation}
\sum_j H_j \Tr \{ H_j Y_k\}  = {N (N+1)\over 2 } Y_k \; ,
\end{equation}
so that the expression for the Hamiltonian generating the projection of $\Lt$ onto $\mathfrak{ham}(\hilbert)$ reads
\begin{equation}
K = -{ \text{i} \over 2N } \sum_k \gamma_k \left[ \Tr \{ E_k \} E_k^{\dagger} - \Tr\{ E_k^{\dagger}\} E_k \right] \; .
\end{equation}
Then, the physical dissipator will be given by the remaining contribution and will have minimal norm:
\begin{align}\nonumber
\Dt[\rho] =& \Lt [\rho]- \Pi(\Lt[\rho])  \\ \nonumber
=& \sum_k \gamma_k E_k \rho E_k^{\dagger} \\
&+ {1\over 2N } \sum_k \gamma_k \left[ \Tr \{ E_k \} E_k^{\dagger} - \Tr\{ E_k^{\dagger}\} E_k, \rho \right] \;.
\end{align}
From the requirement that $\Lt \in \mathfrak{htp}(\hilbert)$, which means that $\Lt$ is not only Hermiticity
but also trace preserving, one gets the additional condition on the set of pseudo-Kraus operators:
\begin{equation}
\sum_k \gamma_k E_k^{\dagger}E_k = 0 \; .
\end{equation}
With this, one can rewrite the dissipator in the desired form
\begin{equation}
\Dt[\rho] =\sum_k \gamma_k \left[ L_k \rho L_k^{\dagger} - {1\over 2} \{ L_k^{\dagger} L_k, \rho \} \right] \; ,
\end{equation}
where the Lindblad operators are found to be traceless:
\begin{equation}
L_k = E_k - {\Tr \{ E_k\} \over N} \mathbb{I} \; .
\end{equation}
Given a generator, fixing the Lindblad operators to be traceless at all times also fixes the expression for the Hamiltonian $K$ (up to a time-dependent constant). Absorption of the rates $\gamma_k$ or a mixing of the Lindblad operators are still allowed, but leave the dissipator and the Hamiltonian separately invariant. Therefore, the dissipator which has minimal norm \eqref{norm} is unique, and is written in terms of traceless Lindblad operators $L_k$.

\bibliography{biblio}

\end{document}